\documentclass[lettersize,journal]{IEEEtran}

\usepackage{amsmath,amsfonts}
\usepackage{algorithmic}
\usepackage{array}
\usepackage[caption=false,font=normalsize,labelfont=sf,textfont=sf]{subfig}
\usepackage{textcomp}
\usepackage{stfloats}
\usepackage{url}
\usepackage{verbatim}
\usepackage{graphicx}
\usepackage{float}
\usepackage{cite}
\usepackage{xcolor}
\usepackage{geometry}
\usepackage{authblk}
\def\BibTeX{{\rm B\kern-.05em{\sc i\kern-.025em b}\kern-.08em
    T\kern-.1667em\lower.7ex\hbox{E}\kern-.125emX}}
\usepackage{balance}

\begin{document}
\title{Activation Map-based Vector Quantization for 360-degree Image Semantic Communication}

\author{Yang Ma\IEEEauthorrefmark{2}, Wenchi Cheng\IEEEauthorrefmark{2}, Jingqing Wang\IEEEauthorrefmark{2}, and Wei Zhang\IEEEauthorrefmark{3}
	\vspace{6pt}
	\\
	
	\IEEEauthorblockA{\IEEEauthorrefmark{2}State Key Laboratory of Integrated Services Networks, Xidian University, Xi'an, China \\
		\IEEEauthorrefmark{3}School of Electrical Engineering and Telecommunications, The University of New South Wales, Sydney, Australia\\
		E-mail: \{\emph{yangma@stu.xidian.edu.cn}, \emph{wccheng@xidian.edu.cn}, \emph{jqwangxd@xidian.edu.cn}, \emph{w.zhang@unsw.edu.au}\}}
\vspace{-20pt}
}

\maketitle
\thispagestyle{empty}
\pagestyle{empty}
\begin{abstract}
In virtual reality (VR) applications, 360-degree images  play a pivotal role in crafting immersive experiences and offering panoramic views, thus improving user Quality of Experience (QoE). However, the voluminous data generated by 360-degree images poses challenges in network storage and bandwidth. To address these challenges, we propose a novel Activation Map-based Vector Quantization (AM-VQ) framework, which is designed to reduce communication overhead for wireless transmission. The proposed AM-VQ scheme uses the Deep Neural Networks (DNNs) with vector quantization (VQ) to extract and compress semantic features. Particularly, the AM-VQ framework utilizes activation map to adaptively quantize semantic features, thus reducing data distortion caused by quantization operation. To further enhance the reconstruction quality of the 360-degree image, adversarial training with a Generative Adversarial Networks (GANs) discriminator is incorporated. Numerical results show that our proposed  AM-VQ scheme achieves better performance than the existing Deep Learning (DL) based coding  and the traditional coding schemes under the same transmission symbols.   

\end{abstract}


\section{Introduction}
\IEEEPARstart{O}mnidirectional image, which is often referred to as 360-degree images, are an emerging media format that provides panoramic views of different scenarios \cite{10277512}. This format enables users to explore environments from various perspectives, thus achieving a comprehensive visual experience. Also, the 360-degree image has gained prominence in the field of virtual reality (VR). In VR applications, 360-degree images and videos serve as the main sources of content. They provide a more realistic and interactive experience by allowing users to be fully immersed in a three-dimensional environment. Compared to traditional planar images, 360-degree images contain a significantly larger amount of content \cite{VR}. However, this has introduced a set of challenges in terms of network storage and bandwidth. Due to their size and complexity, these images require more efficient storage solutions and impose greater demands on network bandwidth. To address these challenges, researchers have explored more efficient image compression technologies and transmission solutions.

Traditional compression techniques for 360-degree images often utilize well-known encoding standards such as JPEG\cite{JPEG}, HEVC\cite{HEVC}, and VP9\cite{MPEG}. The primary goal of these methods is to reduce the file size of image data while maintaining visual quality by exploiting redundancies. Applying traditional compression methods to 360-degree images can be challenging due to their unique characteristics. These characteristics include higher resolution and a wider field of view, resulting in more extensive data than normal images. Storing and transmitting extensive data presents significant challenges in terms of storage and bandwidth. In addition, the irregular geometry and distortions present in 360-degree images require specialized techniques to efficiently manage the data.

Deep learning-based compression methods, which use Deep Neural Networks (DNNs) as their core structures, have shown significant potential in addressing the challenges associated with image compression \cite{deng2021lau}. These methods effectively handle the spatial information of images, capturing both local and global features through a hierarchical approach to multi-scale feature extraction \cite{li2020efficient}. The integration of attention mechanisms and generative models, such as Variational Autoencoders (VAEs) and Generative Adversarial Networks (GANs) \cite{zhang2023gan}, further enhances compression efficiency. In addition, the application of transfer learning and pre-training strategies \cite{Donahue14} shows promise, particularly in scenarios with limited annotated data for 360-degree visuals, facilitating the adaptation of knowledge from traditional image domains and accelerating model convergence for improved compression performance.

At high compression ratios, both traditional techniques and deep learning approaches have limitations in image compression. Traditional methods often result in significant degradation of image quality at high compression rates,  particularly in preserving detail and texture. Moreover, these techniques may inadequately address the spatial distortions inherent in panoramic images\cite{sullivan2012overview}. Deep learning methods confront challenges in sustaining reconstruction quality at extremely high compression ratios. These challenges become more pronounced in scenarios where the models have limited generalization capabilities or where there is a lack of comprehensive training data \cite{jamil2023learning}.

In this paper, we propose an Activation Map-based Vector Quantization (AM-VQ) framework designed for efficient 360-degree image semantic communication with minimal transmission overhead. The AM-VQ framework specializes in extracting and compressing features to minimize the transmitted bit count. More specifically, DNNs are employed to extract multi-scale image features, which are subsequently quantized using the Vector Quantization (VQ) method, resulting in a substantial reduction in the transmission cost of 360-degree images. Moreover, the proposed AM-VQ scheme incorporates an activation map to adaptively quantify semantic features, thereby reducing data distortion caused by quantization.

The rest of this paper is organized as follows. Section II introduces  the VQ semantic communication framework. In Section III, we introduce the AM-VQ 360-degree image semantic communication framework. Then, performance analyses are given in Section IV.  Finally, the  conclusions are drawn in Section V.

\section{SYSTEM MODEL}
\begin{figure}[htbp]
	\centering
	\includegraphics[width=0.5\textwidth]{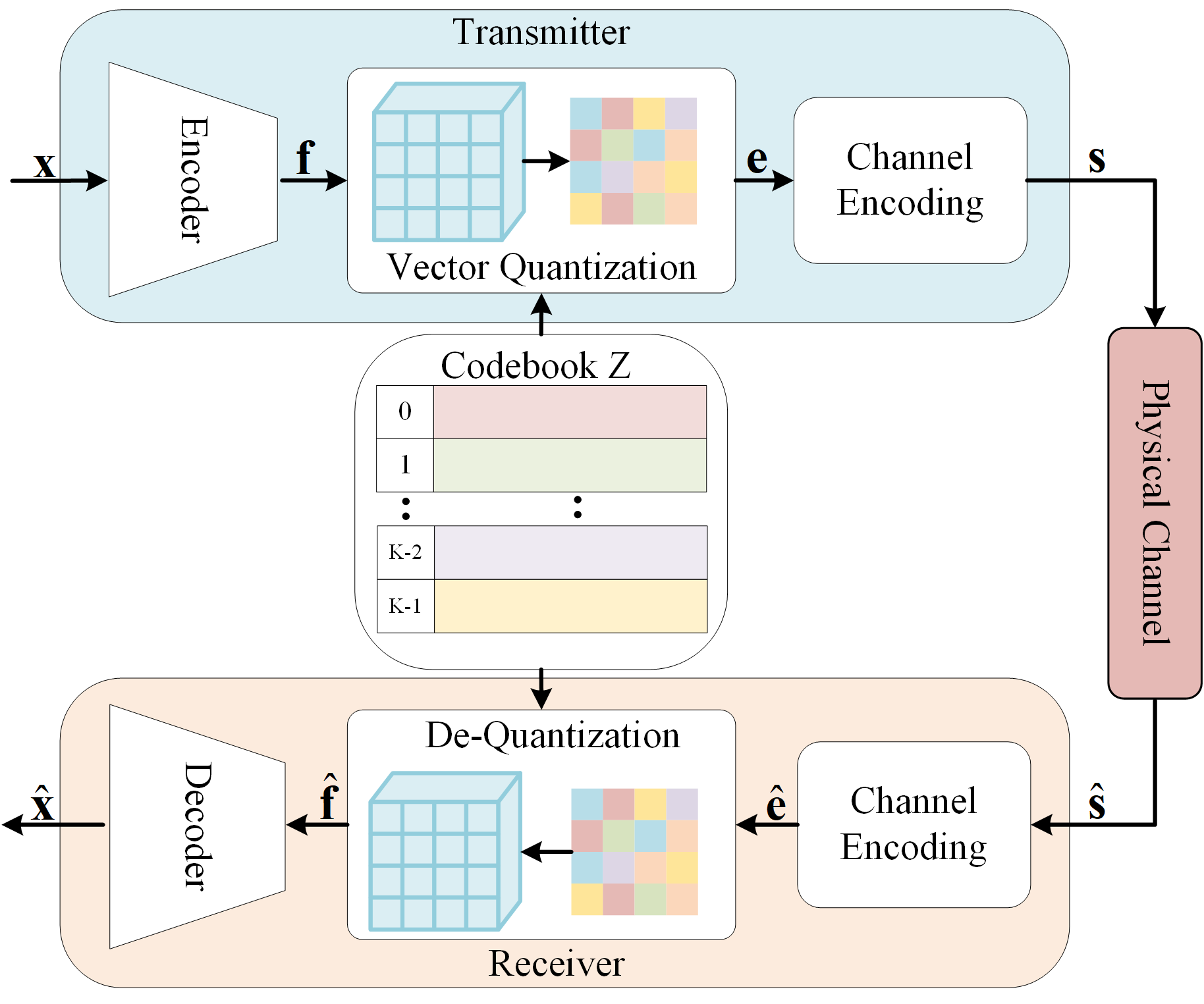}
	\caption{Vector quantization based 360-degree image semantic communication framework.}
	\label{fig1}
\end{figure}

In this section, we consider a Deep Learning-enabled end-to-end 360-degree image semantic communication framework with a physical channel, as shown in Fig. \ref{fig1}.

The transmitter consists of three parts: semantic encoder, VQ module, and channel encoder. We denote the input image as $\bold{x}$. The semantic feature $\bold{f}$ is extracted first as 
\begin{align}
	\bold{f}=E(\bold{x};\theta_{1}),
\end{align}
where $\theta_{1}$ represents the trainable parameters of the semantic encoder $E$, $\bold{f}=\left \{\bold{f}_{1}, \cdots,\bold{f}_{m}, \cdots,\bold{f}_{M}   \right \}\in \mathbb{R}^{M\times L}$, where $M$ represents the number of semantic features, and $L$ is the dimension of each vector. Particularly, a local codebook shared by the receiver is denoted as $\bold{z}=\left \{\bold{z}_{1}, \cdots,\bold{z}_{k}, \cdots,\bold{z}_{K} \right \}\in \mathbb{R}^{K\times L}$, where $K$ represents the number of vectors in the codebook $\bold{z}$. Then with VQ, the semantic feature $\bold{f}_{m}$ can be mapped to an index $k$ as 	 
\begin{align} \label{eq2}
	k = Q(\bold{f}_{m}):=\left(  \arg \min_{\bold{z}_{k}\in \bold{z}}\left \| \bold{f}_{m}-\bold{z}_{k} \right \|  \right ), 
\end{align}
where $k\in [1, K]$ is the index of the mapping $\bold{z}_{k} $ of $\bold{f}_{m} $ into $\bold{z}$. The semantic features can be mapped onto a sequence of indices $\bold{e}$ using VQ. After conventional channel coding, $\bold{e}$ is processed as a transmitted signal $\bold{s}$.

The receiver also consists of three parts: channel decoder, de-VQ, and semantic decoder. The received signal $\hat{\bold{s}}$ can be expressed as
\begin{align}
	\hat{\bold{r}}=\bold{h} \odot \bold{s} + \bold{n},
\end{align} 
where $\bold{h}$ represents the linear channel coefficients between the channel encoder and decoder.  In the context of Rayleigh fading, the channel noise is assumed to follow a normal distribution with mean $0$ and variance $\sigma^2$. 

Assuming the receiver possesses perfect channel state information, the recovered signal is $\hat{\bold {s}}$,
The channel decoder recovers the indices $\hat{\bold{e}}$ from $\hat{\bold{s}}$.  Subsequently, by selecting the corresponding vectors in the codebook with $\hat{\bold{e}}$, the receiver can reconstruct the semantic feature $\hat{\bold{f}}$. Finally, the semantic decoder generates the reconstructed image, which is given by $\hat{\bold{x}}=D(\hat{\bold{f}};\theta_{2})$, where $D$ is the semantic decoder with learnable parameters $\theta_{2}$.

\section{The Design of Proposed Activation Map-based Vector Quantization Semantic Communication Framework}
In this section, we first introduce our proposed AM-VQ framework. Then, we discuss the details of this framework and the loss function.
\subsection{Activation Map-Based Vector Quantization }
VQ is a lossy compression technique that inherently involves discarding original data during the quantization process, leading to irreversible information loss and an inability to fully restore decompressed images to their original state. In VQ, semantic features are replaced with the closest vectors from the codebook, which can result in quantization loss due to mismatches between the codebook vectors and the semantic features. The VQ loss function is defined as  
\begin{align}
	\ell _{VQ}=\sum_{m} ( \left \| sg\left [\bold{f}_{m} \right ] -\bold{z}_{k}  \right \|^{2}_{2}+\beta \left \| sg\left [ \bold{z}_{k} \right ] -\bold{f}_{m}  \right \|^{2}_{2} ), 
\end{align} 
where  $sg\left [ \cdot  \right ] $	represents the stop-gradient operation, and $\left \| sg\left [ \bold{z}_{k} \right ] -\bold{f}_{m}  \right \|^{2}_{2}$ is the commitment loss, and $\beta$ is the weighted factor. The reason why we need to set $sg\left [ \cdot  \right ] $  between $\bold{f}_{m}$ and $\bold{z}_{k}$ is because we have discretized the transformation between these two features, and if the L2 loss is calculated directly, it will cause the neural network gradient not being passed back. The common use of uniform quantization strategies in VQ framework further compounds these errors by treating all image blocks identically, thus failing to consider the diverse characteristics and visual importance of different image regions. The quantization process in VQ, characterized by quantization errors due to high compression scenarios, is exacerbated by uniform quantization strategies that neglect the diversity of image regions. Increasing the quality and size of the codebook can reduce quantization loss, but it can also reduce the compression rate and increase computational complexity.  When attempting to improve the quality and size of the codebook, it is important to consider the trade-off between compression efficiency, image quality, and computational resources.
\begin{figure}[htbp]
	\centering
	\includegraphics[width=3.5in]{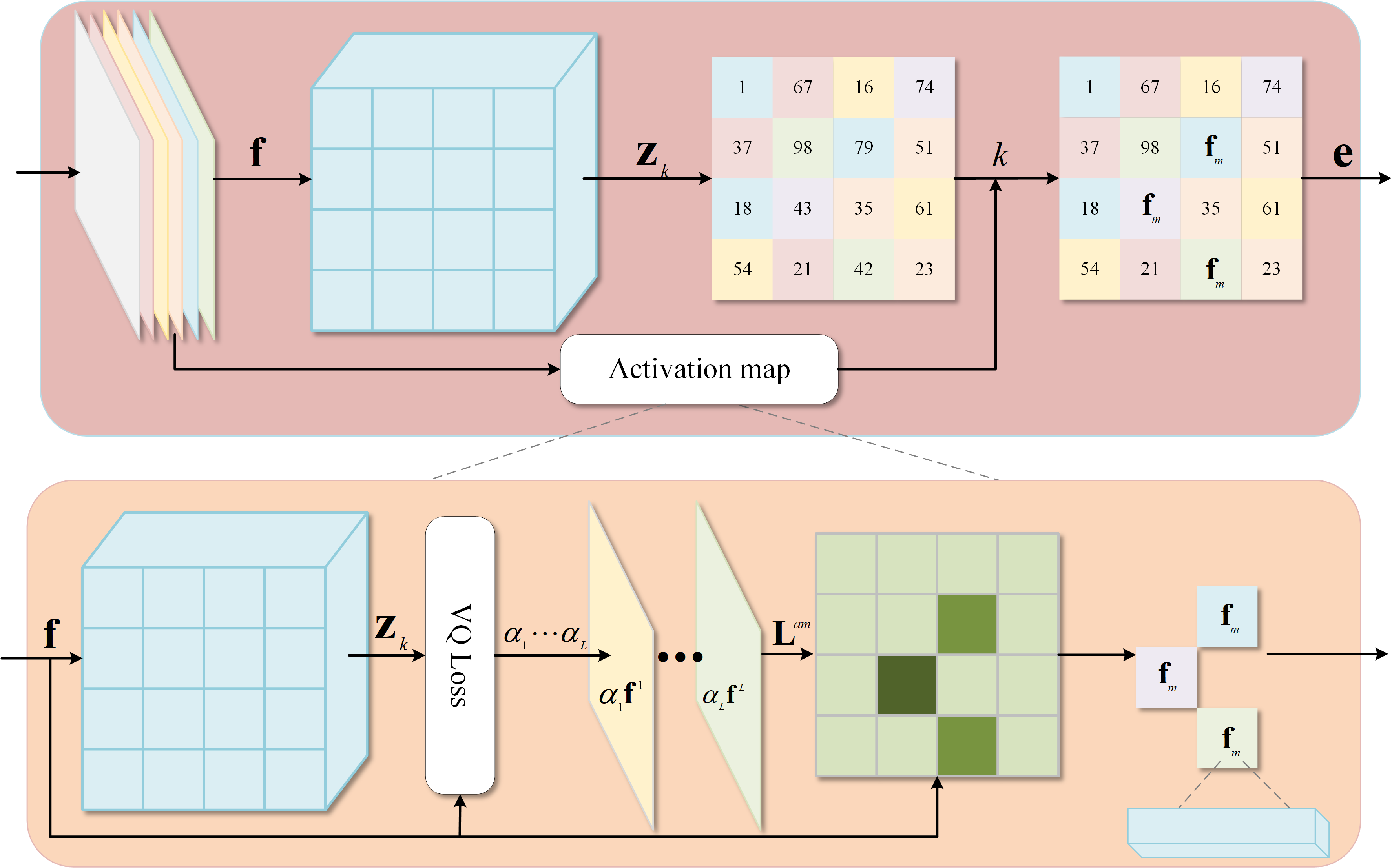}
	\caption{Framework of proposed AM-VQ. }
	\label{fig2}
\end{figure}

The last feature layer encompasses all the semantic information about our target of interest. However, it remains unclear which semantic information undergoes distortion during vector quantization. To address this issue, we propose AM-VQ, whose architecture is shown in Fig. \ref{fig2}. The proposed scheme back-propagates the quantization loss to ascertain the corresponding gradient information of the last feature layer. Global pooling of gradient information produces an activation map for the feature layer, depicting the effect of each feature vector on quantization distortion. The greater the impact, the more significant the distortion of that specific feature vector. Initially, we calculate the gradient from the fully connect $\ell _{VQ}$ network to the  $l-\mathrm{th}$ output feature map $\bold{f}^{l} \in \mathbb{R}^{M\times 1} $ of a convolutional layer. We obtain the weight $\alpha _{l}$ by global average pooling
\begin{align}
	\alpha _{l} =\frac{1}{M}\sum_{m}\frac{\partial \ell _{VQ} }{\partial \bold{f}^{l}_{m}},
\end{align}  
where $m\in [1, M]$. Finally, all the feature maps of final convolutional layer are summed by $\alpha _{l}$ weights and ReLU activation is performed to obtain the activation map, which is referred to as    
\begin{align}
	\bold{L}^{am}=Relu\left(\sum_{l}\alpha_{l}\bold{f}^{l}\right ). 
\end{align} 

A 360-degree image offers a broader range of scenes compared to a conventional image. However, certain regions of such images cannot be accurately characterized after vector quantization. While it is not possible to completely eliminate the distortion caused by the vector quantization, our goal is to reduce the information distortion in the feature maps. To achieve this, we propose a thresholding procedure, in which we do not quantize features in regions where vector quantization characterization would lead to severe distortion. In contrast, features are quantized in the remaining regions. Furthermore, a threshold of $T$ is utilized to filter features by
\begin{align}
	\bold{e}_{m} =\begin{cases}
		Q(\bold{f}_{m}),\quad   &\text{ if } L^{am}_{m}\le T; \\
		\bold{f}_{m},\quad  &\text{ otherwise, } 
		\label{eq:eq9}
	\end{cases}
\end{align}
where $\bold{e}=\left \{\bold{e}_{1}, \cdots,\bold{e}_{m}, \cdots,\bold{e}_{M}\right \}$ result from the fusion of feature vector and quantization vector. The threshold value $T$ is determined during the training process. Thereafter, $\bold{e}_{m}$ undergoes transformation into bit streams $\boldsymbol{s}$, which are then transmitted following channel coding and modulation. In the de-quantization part, $\hat{\bold{f}}$ is recovered by selecting the corresponding vectors from the codebook with the $\hat{\bold{e}}_{m}$. Then, the reconstructed feature tensors of $\hat{\bold{f}}$ are fed into the corresponding layers of the semantic decoder. 

\subsection{Model Description}

\begin{figure*}[htbp]
	\centering
	\includegraphics[width=6.5in]{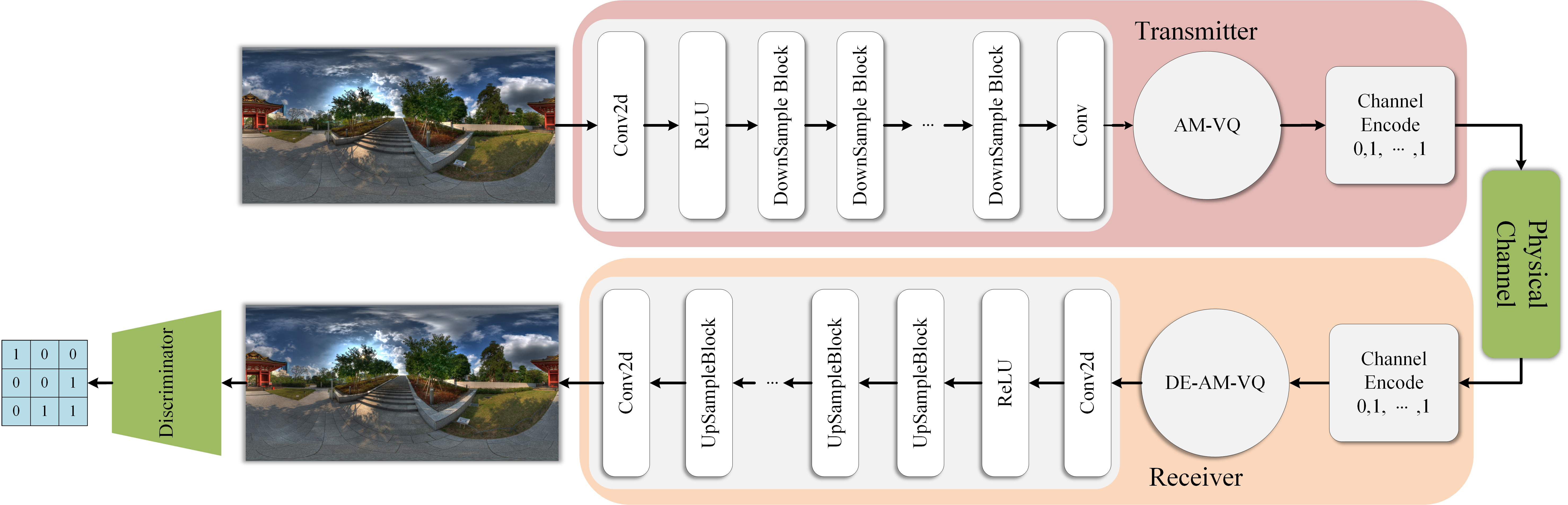}
	\caption{The neural network architecture of the proposed 360-degree image semantic communicate scheme. }
	\label{fig3}
\end{figure*}
The proposed scheme's architecture is illustrated in Fig. \ref{fig3}, which consists of a transmitter, receiver, and physical channel. The transmitter comprises a semantic encoder, AM-VQ, and channel encoder. The semantic encoder consists of a convolutional layer with a ReLU activation function and $L$ downsample blocks that are connected sequentially. Each downsample block connects two residual blocks and a strided convolutional downsample layer. The block concludes with a batch-normalized convolutional layer, after being processed by a ReLU activation function. The strided convolutional downsample layer reduces the dimensions of the feature tensor by subsampling it with a factor of the stride. Afterward, the output feature tensor is fed into the AM-VQ part. The residual block comprises two convolutional layers with batch normalization and ReLU activation function between them.

The receiver consists of a channel decoder, de-AM-VQ , and semantic decoder. The semantic decoder employs a convolutional layer and a ReLU activation function as its input, succeeded by $L$ upsample blocks, and finalized with a convolutional layer. The upsample block is comprised of a sequence connection of a residual block, an upsample layer, and a convolutional layer, followed by batch normalization and ReLU activation function. The upsample layer and the convolutional down-sample layer achieve contrasting effects.

The AM-VQ module, which includes a VQ part and a activation map part, performs quantization encoding and fusion of feature vector and quantization vector. The patch-based discriminator includes a convolutional layer and an activation function as well as three sets of convolutional layers, batch normalization, and activation functions. The design concludes with an additional convolutional layer serving as output. 

\subsection{Loss Function}
The framework is designed to ensure high-quality image transmission while minimizing channel consumption. The training process is divided into two parts: the first part is used for image reconstruction, while the second part focuses on optimizing the codebook. The self-supervised loss during this training process is as
\begin{equation}
	\begin{split}
		\ell _{REC}(E,D,Q)=&  \left \| \bold{x}-\hat{\bold{x}} \right \|^2 +\sum_{m}(\left \| sg\left [  \hat{\bold{f}}_{m} \right ] -\bold{z}_{k}  \right \|^{2}_{2}\\&+ \beta \left \| sg\left [ \bold{z}_{k} \right ] - \hat{\bold{f}}_{m}  \right \|^{2}_{2}) ,
	\end{split}
\end{equation} 
where $\left \| \bold{x}-\hat{\bold{x}} \right \|^2$ is the reconstruction loss. To enhance the visual impact of the image, we employ a PatchGAN discriminator [10] with Binary Cross Entropy (BCE) loss to facilitate adversarial training. The PatchGAN discriminator is capable of evaluating image patches, directing the model to focus more on image details, and its loss function is 
\begin{align}
	\ell_{GAN}(\left \{E,D,Q \right \}, G)= [\log{G\bold{(x)}}+ \log{(1-G\bold{(\hat{x})})}],
\end{align} 
where $G$ is discriminator. The complete objective for finding the optimal compression model then reads
\begin{equation}
	\begin{split}
	\ell  = \arg \min\mathbb{E}_{x\sim p(x)}[ &\ell_{REC}( E,D,Q) -  \\&  \lambda \ell_{GAN}\left \{ ( E,D,Q),G \right \} ]  ,
	\end{split}
\end{equation} 
where $\lambda$ determines the weight that governs the balance of the $\ell_{REC}$ and $\ell_{GAN}$. The appropriate combination of the Least Absolute Deviation (LAD) loss and the GANs loss not only improves the clarity of the generated images, but also ensures that the generated images are more realistic.

\section{NUMERICAL RESULTS}

\subsection{Experiment Setup}

All experiments were conducted on a workstation equipped with 3 Tesla V100 GPUs, utilizing CUDA 11.3 and CuDNN 8.2.1 for parallel acceleration. Based on initial experiments, the embedding space dimension of the codebook is typically set to 1024. A total of 19,859 high-quality 360-degree images were collected from the Flickr photo sharing website and downsampled to mitigate potential compression artifacts\cite{balle2016end}. 
\begin{figure}[H]
	\centering
	\includegraphics[width=3.5in]{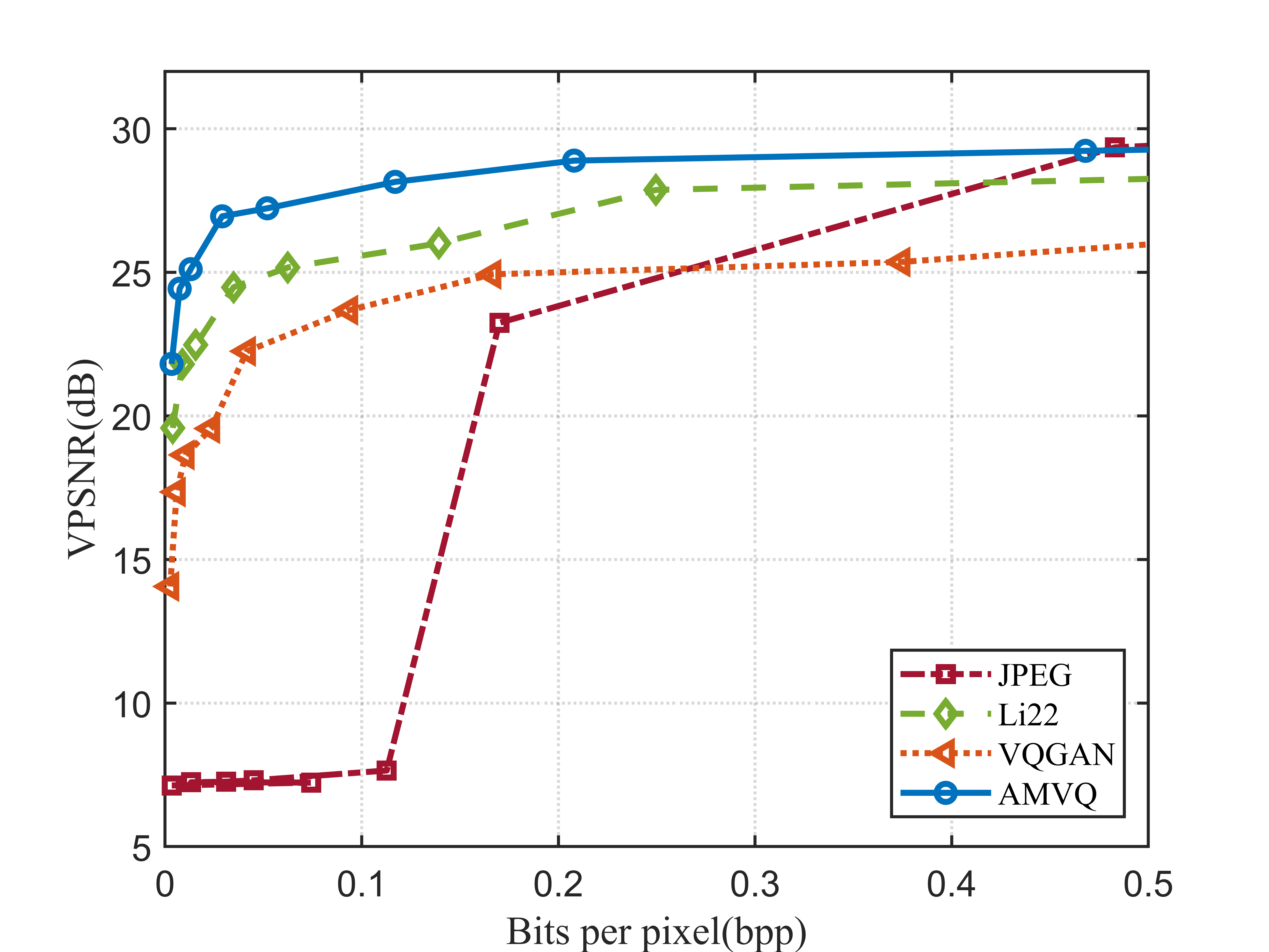}
	\caption{VPSNR performance comparisons versus different compression ratios}
	\label{fig4}
\end{figure}  
The models were trained under noiseless conditions with hyper-parameters $\beta=0.25$ and $\lambda=0.8$. Compression techniques are assessed based on two key aspects: rate and distortion. To estimate the rate, defined as the total number of bits required to encode an image per pixel, the bits per pixel (bpp) method is employed.  As the bpp decreases, the compression rate increases. Distortion is evaluated using viewpoint-based metrics, including Viewpoint-based Peak Signal-to-Noise Ratio (VPSNR) and Viewpoint-based Structural Similarity Index (VSSIM), as indicated by our experiments. The aforementioned metrics, commonly utilized in conventional image algorithms for gauging image reconstruction quality, focus on comparing errors in pixel distances. Consequently, these metrics are unsuitable for assessing the semantic quality of images generated by deep networks. Therefore, perceptual loss as an assessment metric \cite{wang2018high},  which is determined by evaluating the difference between the original image and the image features derived from the restored image using conventional deep neural networks such as VGG \cite{simonyan2014very}, where the perceptual loss is
\begin{align}
P(\bold{x},\hat{\bold{x}})=\left \| \mathrm{conv3}(\bold{x})- \mathrm{conv3}(\hat{\bold{x}})\right \|^{2},
\end{align} 
where $\mathrm{conv3}$ represents the output features of the third layer of the convolutional module. 
\begin{figure}[H]
	\centering
	\includegraphics[width=3.5in]{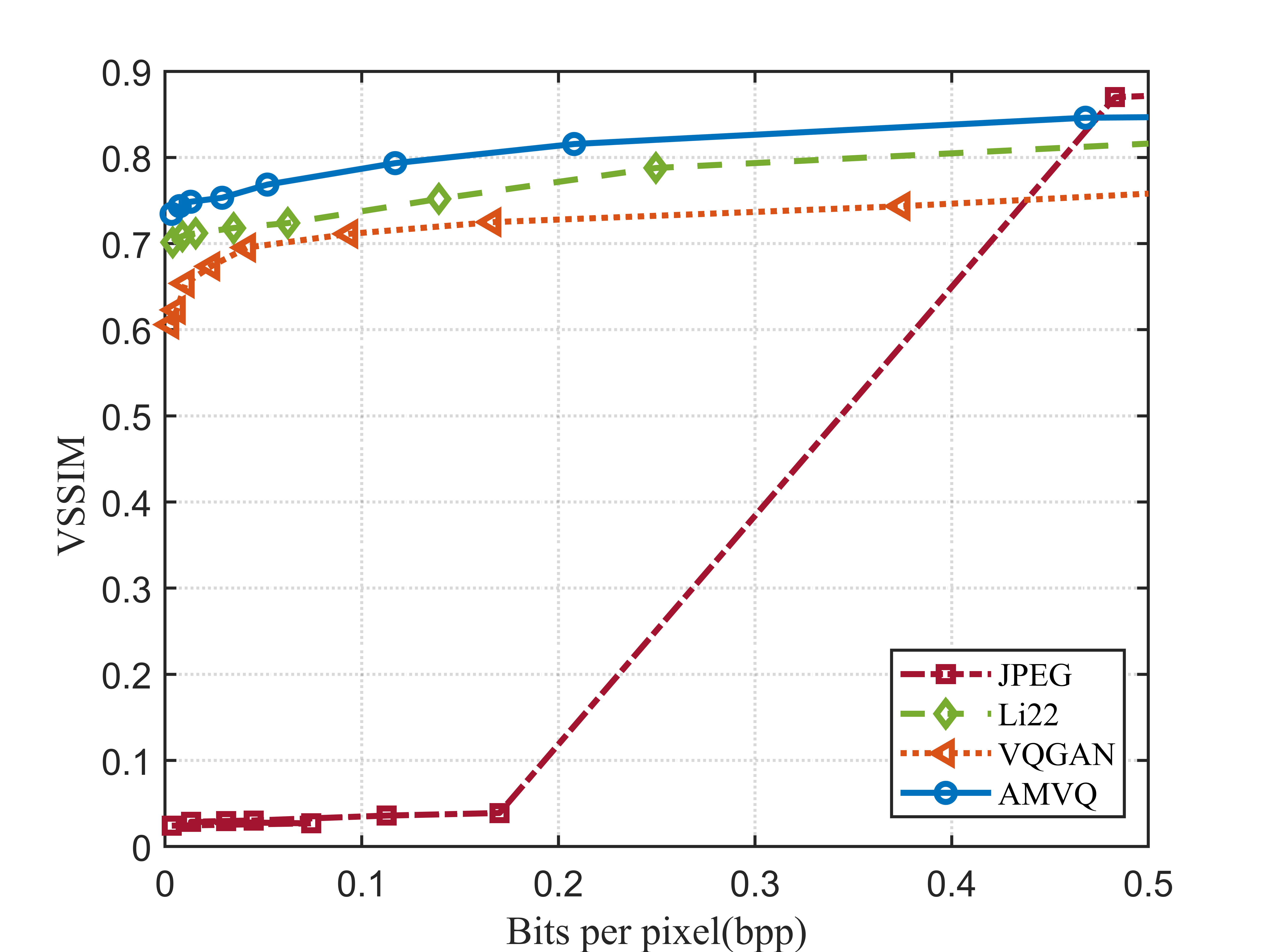}
	\caption{VSSIM performance comparisons versus different compression ratios.}
	\label{fig5}
\end{figure} 

\subsection{Visual Quality Evaluation}
\begin{figure}[H]
	\centering
	\includegraphics[width=3.5in]{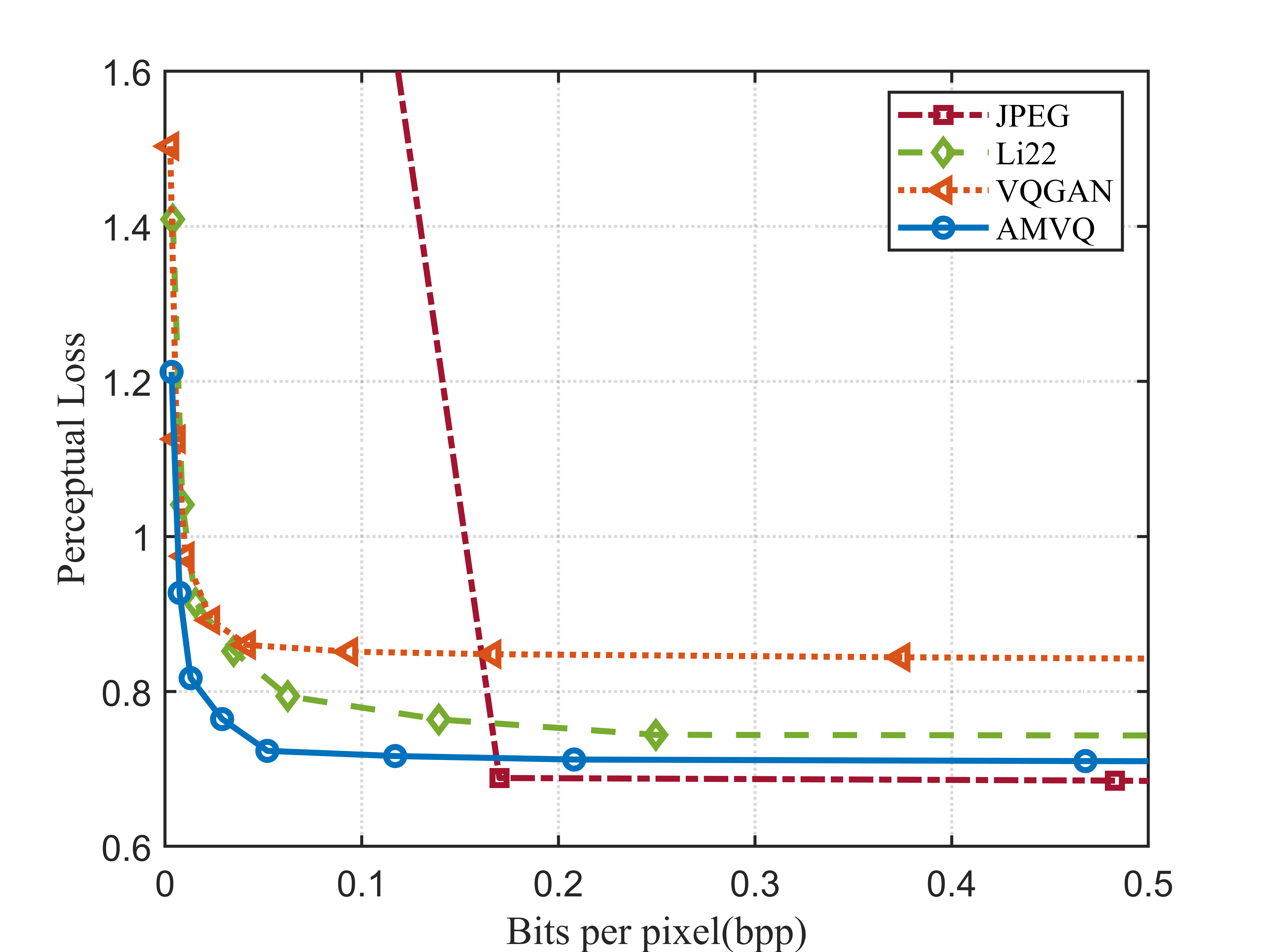}
	\caption{Perceptual Loss performance comparisons versus different compression ratios.}
	\label{fig6}
\end{figure}
We compare our method to traditional codes, i.e., JPEG  and two learned image compression methods, i.e., Li22 \cite{li2022end}, VQGAN \cite{VQGAN}. Fig. \ref{fig4} shows a comparison of the VPSNR image reconstruction effectiveness for various compression ratios. These results demonstrate that the proposed algorithm is superior to conventional methods and other deep learning approaches. Furthermore, the use of deep learning technology for transmitting compressed images does not result in a significant reduction in performance due to the ‘cliff effect’. This effect occurs when the compression ratio exceed a certain threshold, making it impossible for the receiver to retrieve the transmitted image. In contrast, traditional JPEG settings produce noticeable distortions in images when higher compression ratios are applied. However, the deep learning approach can accurately reconstruct the targeted information. The compression transmission scheme we propose in this paper enhances image information representation while conserving bandwidth during signal compression and reconstruction. As compression ratios raise, the performance gap between our suggested algorithm and other models widens. In low compression environments, our algorithm performs comparably to the Mu22 model but is less effective than JPEG. This performance disparity arises because the codebook distorts image details during feature quantization under these conditions, leading to inaccuracies in the detailed information within the reconstructed target.

\begin{figure}[H]
	\centering
	\includegraphics[width=3.5in]{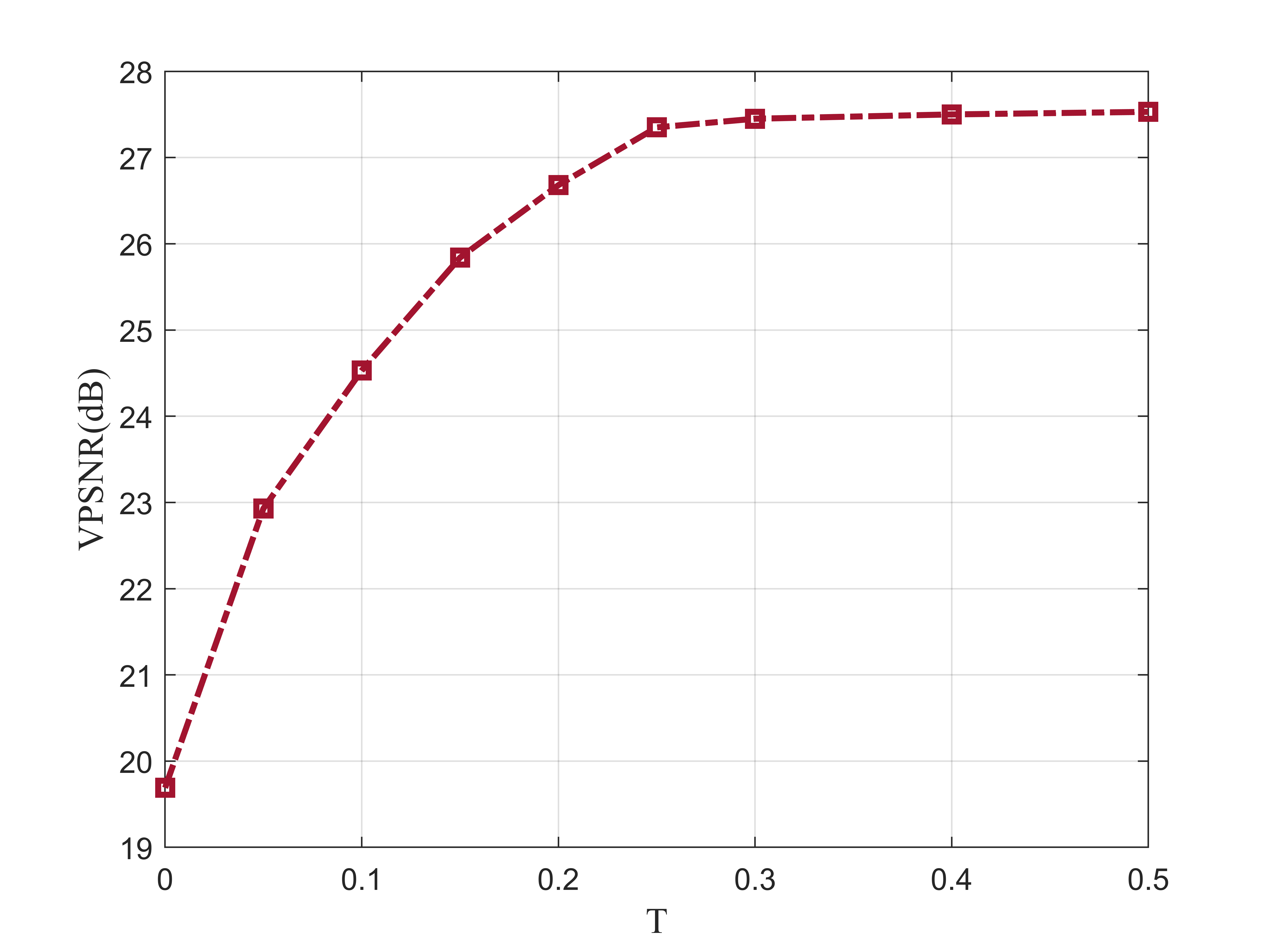}
	\caption{VPSNR of reconstructed images under different threshold $T$.}
	\label{fig7}
\end{figure}
 
Figure. \ref{fig5} also shows the performance comparison of the different methods under the VSSIM evaluation criterion, which reflects the similarity between the original and reconstructed images in terms of image structure. The proposed scheme significantly outperforms the other schemes at high compression ratios, even without using SSIM as the loss function.. The image reconstruction module of the proposed method deeply mines the high-level information of the image and uses the residual network to learn the fused multilevel information, thus enhancing the quality of the reconstructed image.  At high compression ratios, VQGAN shows the poorest reconstruction quality due to distortion from feature vectorization. Conversely, our proposed scheme uses low-latency features for distorted regions through AM, ensuring detailed restoration at these ratios and highlighting our method’s bandwidth efficiency.

Figure \ref{fig6} displays the results of our experiments using the perceptual loss quality metric. Our method demonstrated slight improvements over the comparison methods at low bit rates and clear superiority at high bit rates, compared to state-of-the-art techniques. At low ratios, JPEG can be approximated as lossless compression, which outperforms deep learning-based generation methods. The experimental results indicate that our proposed approach outperforms other methods in vision-related aspects.
\begin{figure}[H]
	\centering
	\includegraphics[width=3.5in]{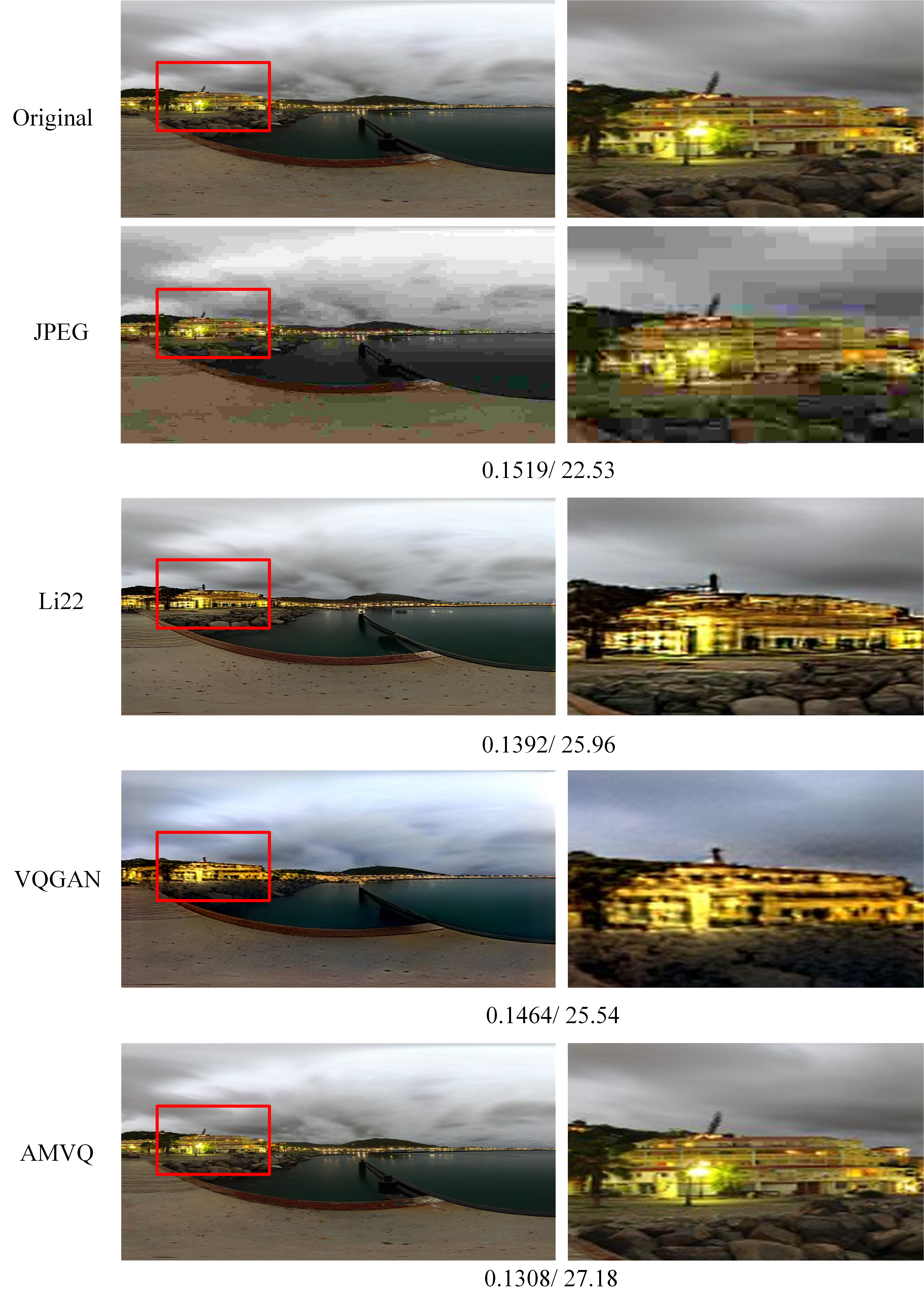}
	\caption{Example images comparing our best-model results with JPEG, Li22, VQGAN. We provide the distortion in form of bpp/VPSNR (dB) under each image. We zoom in some detail regions(as shown in red box).}
	\label{fig9}
\end{figure} 

Figure \ref{fig7} illustrates the VPSNR after image reconstruction with varying threshold values of $T$. When $T$ is set to 0, the VPSNR reaches its minimum value. As the threshold value increases, additional image details are incorporated into the quantized features, leading to a rise in VPSNR. In practical experiments, to achieve a balance between image distortion and visual quality, $T$ is optimally set at 0.3. 

To mitigate the limitations inherent in objective evaluation indices, compression results from various techniques are presented for subjective evaluation. This method facilitates a visual representation of image quality. A 360-degree image has been selected for quality comparison with the original, JPEG, Li22, VQGAN, and our proposed method. The visualization outcomes of these 360-degree images, utilizing various methods, are depicted in Fig. \ref{fig9}. The images have been analysed in greater detail by delineating the focused regions with boxes and extracting and magnifying the corresponding areas. This allows for a more intuitive comparison of the various methods. Furthermore, it can be observed that our method exhibits a clearer texture, which better retains the image details.

\section{CONCLUSION}
In this paper, we propose a framework for AM-VQ semantic communication. The AM-VQ framework combines deep neural networks with VQ to extract and compress semantic features, which are then adaptively quantized using activation maps. Specifically, the AM-VQ technique is used to compress a portion of semantic feature vectors into a corresponding set of semantic feature indexes, which reduces the transmission channel consumption and ensures the transmission quality of 360 images at the same time. Additionally, adversarial training is used to enhance the quality of received images with a PatchGAN discriminator. Numerical results show that our proposed AM-VQ scheme can achieve better performance compared with Deep codec scheme and traditional codec scheme in the 360-degree transmission task.

\bibliographystyle{IEEEtran}
\bibliography{ref_globecom.bib}

\end{document}